\documentclass{article}
\usepackage{fleqn}
\usepackage{epsfig}
\usepackage{amsmath,amssymb}
\begin{document}

\begin{center}
{\bf\Large Nonminimal Macroscopic Models of a Scalar Field Based on Microscopic Dynamics. II. Transport Equations}\\[12pt]
Yu.G. Ignat'ev\\
Kazan Federal University,\\ Kremlyovskaya str., 35,
Kazan 420008, Russia
\end{center}

{\bf keywords}: Relativistic Kinetics, Phantom Scalar Fields,
Scalar Interaction of Particles, Negatives Masses.

\begin{abstract}
The article proposes generalizations of the macroscopic model of plasma of scalar charged particles to the cases of inter-particle interaction with multiple scalar fields and negative effective masses of these particles. The model is based on the microscopic dynamics of a particle at presence of scalar fields. The theory is managed to be generalized naturally having strictly reviewed a series of its key positions depending on a sign of particle masses. Thereby, it is possible to remove the artificial restriction contradicting the more fundamental principle of action functional additivity.
Additionally, as a condition of internal consistency of the theory, particle effective mass function is found.
\end{abstract}
\newcommand{\beqdis}[2]{\begin{equation}\label{#1}
{\displaystyle #2} \end{equation}}
\newcommand{\dsp}{\displaystyle}
\newcommand{\Req}[1]{(\ref{#1})}
\newcommand{\krugskob}[1]{\left(#1\right)}
\newcommand{\kvadrskob}[1]{\left[#1\right]}
\newcommand{\figurskob}[1]{\left\{#1\right\}}
\newcommand{\Pg}{{\rm P}}
\newcommand{\PAcontr}[2]{\Pg^{#1} - e_{#2}A^{#1}}
\newcommand{\Lee}[1]{\stackunder{#1}{\rm L}}
\newcommand{\dx}[1]{\partial_{#1}}
\newcommand{\qf}[1]{\displaystyle 1 + \frac{q_{#1}\Phi}{m_{#1}}}
\newcommand{\Eps}{{\cal E}}
\newcommand{\Const}{\mathop{\rm Const}\nolimits}
\newcommand{\Ham}{{\cal H}}
\newcommand{\dfeq}{\stackrel{def}{=}}
\newcommand{\xsort}[2]{\stackunder{#1}{#2}}
\newcommand{\onetwo}{\frac{1}{2}}
\newcommand{\dual}[1]{\stackrel{*}{#1}}
\newcommand{\anti}[1]{\overline{#1} \,}
\newcommand{\dpox}[2]{\displaystyle \frac{\partial #1}{\partial
#2}}
\newcommand{\xsorancov}[3]{\xsort{#1}{\anti{#2}}\,_{#3}}
\newcommand{\intx}[1]{\displaystyle \int\limits_{#1}^{}}
\newcommand{\sumx}[1]{\displaystyle \sum\limits_{#1}^{}}
\newcommand{\Puass}[1]{\left[\Ham\,, #1\right]}
\newcommand{\llsim}{\stackrel{<}{\sim}}
\newcommand{\noneq}[1]{\begin{equation} #1 \nonumber\end{equation}}
\def\stackunder#1#2{\mathrel{\mathop{#2}\limits_{#1}}}

This work was founded by the subsidy allocated to Kazan Federal University for the state assignment in the sphere of scientific activities.

\section{Introduction}
\label{intro}

In the previous article \cite{Ignatev_print} the Author considered certain strict consequences of the dynamic equations for the statistical systems of
scalar charged particles without the suggestion of nonnegativity of the effective rest mass of the scalar charged particles. The removal of the restriction on the sign of the effective mass of  scalar charged particle allows us to make the dynamic theory compatible with the action additivity principle. In the article we build the complete system of macroscopic equations for the statistical system of scalar charged particles without a restriction on the sign of the effective mass. Also we specify the form of the mass function.

\section{The Rate of Change of the Dynamic Averages}

Let us now calculate the rate of change of the dynamic function's average $\psi(s)$  \cite{Ignatev_print}
\begin{equation}\label{Psi(tau)dG}
\Psi(\tau)=
\int\limits_\Omega F(\eta(s))\psi(\eta(s))\delta(s-s(\tau))d\Gamma.
\end{equation}
Calculating the time $\tau$ derivative of both parts of (\ref{Psi(tau)dG}) with an account of the symbolic rule for the differentiation of
Dirac $\delta$ - function
$$\frac{d}{dx}\delta(g(x))=\delta(g(x))\frac{d}{dx},$$
we fin:
\begin{eqnarray}\label{psi_dt}
\frac{d\Psi(\tau)}{d\tau}=& \displaystyle\int\limits_\Omega \delta(s-s(\tau))\times\nonumber\\
& \displaystyle\frac{ds}{d\tau}\frac{d}{ds}\biggl(F(\eta(s))\psi(\eta(s))d\Gamma\biggr).
\end{eqnarray}
In the integral (\ref{psi_dt}) we take into account the constancy of the particle phase space
 \cite{Ignatev_print} $d\Gamma/ds=0$ and the relation for the total derivative of the dynamic function
\begin{equation} \label{Eq2}
\frac{d\Psi }{ds} =[H,\Psi ].
\end{equation}
Then we obtain:
\begin{eqnarray}
\frac{d\Psi(\tau)}{d\tau}=
{\displaystyle\int\limits_\Omega \frac{d}{ds}}
{\displaystyle\biggl(F(\eta(s))\psi(\eta(s))\biggr)\times}\nonumber\\
\delta(s-s(\tau))\frac{ds}{d\tau}d\Gamma
= {\displaystyle\int\limits_\Omega [H,F\psi]\delta(s-s(\tau))\frac{ds}{d\tau}d\Gamma}\nonumber
\end{eqnarray}
Let us then take into account relation \cite{Ignatev_print}
\begin{equation} \label{Eq4}
\frac{dH}{ds} =[H,H]=0,\Rightarrow H= \Const.
\end{equation}
and the linearity of the Poisson bracket:
\begin{equation}\label{HF}
[H,F\psi]=[H,f\delta(H)\psi]=\delta(H)[H,f\psi].
\end{equation}
Then carrying out the integration over time derivative and mass surface, we find:
\begin{equation}\label{dPsi_tau}
\frac{d\Psi(\tau)}{d\tau}= {\displaystyle\int\limits_{\Omega_0} m_*[H,f\psi]d\Gamma_0}
\end{equation}
In particular, for the particle number's rate of change
\begin{equation}\label{dN}
dN(\tau)=F(x,P)\delta(s-s(\tau))d\Gamma
\end{equation}
putting $\psi=1$ in (\ref{dPsi_tau}), we find:
\begin{equation}\label{dNdtau}
\frac{dN(\tau)}{d\tau}= \int\limits_{\Omega_0} m_*[H,f]d\Gamma_0
\end{equation}
Let us now take into account the differential equation \cite{Ignatev_print} for the Hamilton function
of scalar charged particles
\begin{equation}\label{HPsi}
[H,\Psi]=\frac{1}{m_*}P^i\widetilde{\nabla}_i\Psi+\partial_i m_*\frac{\partial \Psi}{\partial P_i}.
\end{equation}
Then we finally obtain:
\begin{equation}\label{dPsi-tau1}
\frac{d\Psi(\tau)}{d\tau}= {\displaystyle\int\limits_{\Omega_0} \biggl(%
P^i\widetilde{\nabla}_i +\frac{1}{2}\partial_i m_*^2\frac{\partial}{\partial P_i} %
\biggr)f\psi d\Gamma_0}.
\end{equation}
Thus, the relation for the change rate of the dynamic averages does not depend on the sign of a mass function.

Let now range $\Omega_0$ covers the entire 6-dimensional phase space $\Gamma_0$.
For the sake of simplicity of the first integral in \eqref{dPsi-tau1} it is required to make use of the integral
relation for the Cartan derivative (see e.g., \cite{Ignatev1}):
\begin{equation}\label{int_cart}
\int\limits_{P(X)}\widetilde{\nabla}_i\psi(x,P)dP\equiv \nabla_i\int\limits_{P(X)}\psi(x,P)dP.
\end{equation}
Then let us accept the suggestion regarding features of the dynamic functions
on the infinite sphere $\Sigma_P(X)$ covering 3-dimensional momentum space:
\begin{equation}\label{f8}
\left.f(x,P)\psi(x,P)\right|_{\Sigma_P(X)}\to 0.
\end{equation}
Next, carrying out partial integration, we obtain the following integral relation:
\begin{equation}\label{int_HFPsi}
\int\limits_{P_0(X)}\frac{\partial}{\partial P_i}f(x,P)\psi(x,P)dP_0=0.
\end{equation}
Thus for \eqref{dPsi-tau1} we find:
\begin{equation}\label{nabla_psi}
\frac{d\Psi(\tau)}{d\tau}= {\displaystyle \int\limits_{V}dV\nabla_i\int\limits_{P_0}
P^i f\psi dP_0}.
\end{equation}

\section{Kinetic And Transport Equations}
\subsection{The General Relativistic Kinetic Equations}
Since the factor of the effective mass sign does not explicitly depend on the form of the invariant general - relativistic kinetic equations we confine ourselves to brief information about relativistic kinetic equations (see e.g., \cite{Ignatev3,Ignatev4,Yubook1}). Due to the local conformity principle and the assumption about 4-dimensional pointness of particle collisions, the generalized momentum of the
 system of interacting particles is conserved in each act of interparticle interaction:
\begin{equation} \label{GrindEQ__49_}
\sum _{I} P_{i} =\sum _{F} P'_{i} ,
\end{equation}
where the summation is carried out by all the initial states $P_{i} $ and
final states $P'_{i} $. Let the following reactions run in plasma:
\begin{equation} \label{GrindEQ__50_}
\sum _{A=1}^{m} \nu _{A} a_{A} {\rm \rightleftarrows }\sum _{B=1}^{m'} \nu '_{B} a'_{B} ,
\end{equation}
where $a_{A} $ are particle symbols and $\nu _{A} $ are particle numbers in each channel of reactions. Thus the generalized momentums of the initial and final states are equal:
\begin{equation}\label{GrindEQ__51_}
P_I=\sum\limits_{A=1}^m\sum\limits_\alpha^{\nu_A} P^\alpha_A,
\quad P_F=\sum\limits_{B=1}^{m'}\sum\limits_{\alpha'}^{\nu'_B} P'\
\!\!^{\alpha'}_B.
\end{equation}
The distribution functions of particles are determined by the invariant kinetic equations
 \cite{Ignatev3}\footnote{Normalization factor $m_*$ in the left part of \eqref{GrindEQ__51_} takes
 account of the Hamilton function normalization \eqref{HPsi}.}:
\begin{equation} \label{GrindEQ__52_}
m_*[H_a,f_a]=\mathrm{I}_a(x,P),
\end{equation}
where $\mathrm{I}_{a} (x,P_{a} )$ is an integral of collisions:
\begin{eqnarray} \label{GrindEQ__53_}
\mathrm{I}_{a} (x,P_{a} )=-\sum  \nu _{A} \times\nonumber\\
\int  '_{a} \delta ^{4} (P_{F} -P_{I} )W_{IF} (Z_{IF} -Z_{FI} )\prod _{I,F} 'dP;
\end{eqnarray}
where
$$W_{FI} =(2\pi )^{4} |M_{IF} |^{2} 2^{-\sum  \nu _{A} +\sum  \nu '_{b} } $$
is a scattering matrix of the reaction channel
 \eqref{GrindEQ__50_}, ($|M_{IF}|$ are invariant scattering amplitudes); $I$ is the initial state, $F$ is the final state;
\begin{eqnarray}
Z_{IF} =\prod _{I} f(P_{A}^{\alpha } )\prod _{F} [1\pm f(P_{B}^{\alpha '} )];\nonumber\\
\quad Z_{FI} =\prod _{I} [1\pm f(P_{A}^{\alpha } )]\prod _{F}
f(P_{B}^{\alpha '} ),\nonumber
\end{eqnarray}
the sign ``+'' corresponds to bosons and ``-'' correponds to fermions (please see details in \cite{Ignatev3,Ignatev4}).
\subsection{The Transport Equations of the Dynamic Quantities}
Let us now proceed to deriving the transport equations of the dynamic quantities. %
Using the linearity of the Poisson bracket and making equal the right parts of equations \eqref{dPsi_tau} and
 \eqref{nabla_psi}, we obtain the following integral relation:
\begin{eqnarray}
\displaystyle \int\limits_V dV\biggl(\nabla_i\int\limits_{P_0}P^i f\psi dP_0-\nonumber\\
\int\limits_{P_0}\psi m_*[H,f]dP_0-\int\limits_{P_0}f m_*[H,\psi]dP_0\biggr)=0.
\end{eqnarray}
In consequence of the arbitrariness of range $V$ we hence obtain an integral-differential relation:
\begin{eqnarray}\label{equiv_psi}
\displaystyle \nabla_i\int\limits_{P_0}P^i f\psi dP_0-\nonumber\\
\int\limits_{P_0}\psi m_*[H,f]dP_0-\int\limits_{P_0}f m_*[H,\psi]dP_0=0.
\end{eqnarray}
Substituting instead of the Poisson bracket its
exp\-res\-sion from the kinetic equations \eqref{GrindEQ__52_}
into the second integral in (\ref{equiv_psi}), we obtain:
\begin{equation}\label{equiv_psi1}
\nabla_i\int\limits_{P_0}P^i f\psi dP_0-\int\limits_{P_0}\psi \mathrm{I}_a dP_0-\int\limits_{P_0}f m_*[H,\psi]dP_0=0.
\end{equation}
Summarizing now \eqref{equiv_psi1} by all particle sorts and taking into account the expression for the integral of collisions, we obtain the {\it transport equations of the dynamic quantities} $\psi _{a} (x,P_{a} )$ in the capacity of the general-relativistic kinetic equations \eqref{GrindEQ__52_}:
\begin{eqnarray}
\nabla _{i} \sum _{a} \int\limits _{P_0} \Psi _{a} f_{a}
P^i dP_{a} -\nonumber\\
\sum _{a} \int\limits _{P_0} f_{a} m_*[H_{a} ,\Psi _{a} ]dP_{a} =\nonumber\\
\label{GrindEQ__54_}
-\sum _{by\; chanels} \int  \biggl(\sum _{A=1}^{m} \nu _{A} \Psi _{A} -
\sum _{B=1}^{m'} \nu '_{B} \Psi '_{B} \biggr)\times\nonumber\\
\delta ^{4} (P_{F} -P_{I} )(Z_{IF} W_{IF} -Z_{FI} W_{FI} )\prod _{I,F} dP,
\end{eqnarray}
where the summation is carried out by all the reaction channels
\eqref{GrindEQ__50_}.

Putting $\Psi _{a}
=g_{a} $ in  \eqref{GrindEQ__54_}, where $g_{a} $ are certain fundamental charges being conserved in  reactions \eqref{GrindEQ__50_}, with account of \eqref{GrindEQ__49_}, \eqref{GrindEQ__51_} and \eqref{GrindEQ__54_} we obtain the transport equations of the plasma particle\\ number flux densities:
\begin{equation} \label{GrindEQ__55_}
\nabla _{i} J_{G}^{i} =0,
\end{equation}
where:
\begin{equation} \label{GrindEQ__56_}
J_{G}^{i} =\sum _{a} \frac{2S+1}{(2\pi )^{3} } \; g_{a}
\int\limits_{P_0} f_a(x,P)P^{i} dP_0.
\end{equation}
is a density vector of the fundamental current corresponding to charges $g_{a} $. Particularly, the con\-ser\-vation law
\eqref{GrindEQ__55_} always is held for each particle sort
$b$ ($g_{a} =\delta _{a}^{b} $) given their collisions are elastic.

Let us put  $\Psi _{a} =P^{k} $ in \eqref{GrindEQ__54_} $\Psi _{a} =P^{k} $. Then as a result of the conservation law of the generalized momentum at collisions \eqref{GrindEQ__49_}, the integrand in big parentheses
\eqref{GrindEQ__54_} is equal to:
\begin{equation}
\sum _{A=1}^{m} \nu _{A} \Psi _{A} -
\sum _{B=1}^{m'} \nu '_{B} \Psi '_{B}\equiv P_I-P_F=0.\nonumber
\end{equation}
Thus, taking into account \cite{Ignatev_print}
\begin{equation} \label{Eq9}
[H,P^{k} ]=\nabla ^{k} \varphi \equiv g^{ik} \partial _{i} \varphi;
\end{equation}
and \eqref{GrindEQ__49_} we obtain the transport equations of plasma energy-momentum:
\begin{equation} \label{GrindEQ__57_}
\nabla _{k} T_{p}^{ik} -\sum\limits_r\sigma_{(r)}\nabla ^{i} \Phi_{r} =0,
\end{equation}
where there are introduced the \textit{plasma energy-momentum tensor}
\begin{equation}\label{Tpl}
T^{ik}_p=\sum\limits_{a} \frac{2S+1}{(2\pi )^{3}}
\int\limits_{P_0} f_a(x,P)P^iP^kdP_0
\end{equation}
and the \textit{scalar densities of plasma charge relative to scalar field $\Phi_r$}, $\sigma^{(r)}$ :
\begin{equation}\label{sr}
\sigma^{(r)}=\sum\limits_a \sigma^{(r)}_a,
\end{equation}
where $\sigma^{(r)}_a$ are the \textit{scalar charge densities of $a$-component of plasma relative to scalar field $\Phi_r$}:
\begin{equation} \label{GrindEQ__58_}
\sigma^{(r)}_a =\frac{2S+1}{(2\pi )^{3} } m^*_a q^{(r)}_a
\int\limits_{P_0} f_a(x,P)dP_0,
\end{equation}

Particularly, for charge singlet $(q,\Phi)$
the con\-ser\-vation law \eqref{GrindEQ__57_} takes form:
\begin{equation} \label{GrindEQ__57_0_}
\nabla _{k} T_{p}^{ik} -\sigma \nabla ^{i} \Phi =0,
\end{equation}
where it is (see \cite{Ignatev3,Yu_stfi14}):
\begin{equation} \label{GrindEQ__59_}
\sigma =\Phi \frac{2S+1}{(2\pi )^{3} } q^{2}
\int\limits_{P_0)} f(x,P)dP_0.
\end{equation}
It should be noted that the form of the energy-momentum tensor \eqref{Tpl} and charge scalar density
\eqref{GrindEQ__58_} which was found for scalar charged particles at given Hamilton function, is a direct consequence of the canonical equations and the assumption
about conservation of total momentum in local collisions of particles.
\subsection{Conservation of the Total Energy-Momentum Tensor}
The complete system of macroscopic equations consists, first of all, from the Einstein equations:
\begin{equation}\label{Einst_Scalar}
R^{ik}-\frac{1}{2}Rg^{ik}=8\pi (T^{ik}_p+T^{ik}_s),
\end{equation}
where $T^{ik}_p$ is the determined earlier energy - mo\-mentum tensor of the statistical system and $T^{ik}_s$ is the energy - momentum tensor  of the system of  $N$ inde\-pendent scalar fields:
\begin{eqnarray}\label{Tik_s}
T_{s}^{ik} =\sum\limits_r\frac{\epsilon_1^{(r)}}{8\pi } \biggl[2\Phi_{(r)} ^{,i} \Phi_{(r)} ^{,k}\nonumber\\
 -g^{ik} \Phi_{(r),j} \Phi_{(r)} ^{,j} +
\epsilon^{(r)}_2 {m^{(r)}_s}\!\ ^2 g^{ik} \Phi_{(r)} ^{2} \biggr],
\end{eqnarray}
where for the classical scalar field it is $\epsilon_2=1$, for the fantom scalar field it is  %
$\epsilon_2=-1$; for the field with repulsion of the like charged particles it is $\varepsilon_1=-1$. Let us note that the energy-momentum tensor
of the scalar field in form (\ref{Tik_s}) is obtained from Lagrangian \cite{YuNewScalar3}:
\begin{equation}\label{Ls}
L_s= \sum\limits_r\frac{\epsilon^{(r)}_1}{8\pi}(\Phi_{(r),i}\Phi_{(r)}^{,i}-\epsilon^{(r)}_2 {m^{(r)}_s}\!\ ^2\Phi_{(r)}^2).
\end{equation}
Let us calculate the covariant derivative $\nabla_k$ of the total energy-momentum tensor
\begin{equation}\label{Tik_tot}
T^{ik}=T^{ik}_p+T^{ik}_s.
\end{equation}
Then let us find with account of (\ref{GrindEQ__57_0_}) and (\ref{Tik_s}):
\begin{eqnarray}\label{nabla_Tik}
\nabla_k T_i^k=\frac{1}{4\pi}\sum\limits_r\biggl[\epsilon_1^{(r)}\biggl(\Box \Phi_{(r)}+\nonumber\\
\epsilon^{(r)}_2{m^{(r)}_s}\!\ ^2\Phi_{(r)}\biggr)
+4\pi\sigma^{(r)}\biggr]\nabla_i\Phi_{(r)}=0.
\end{eqnarray}
Due to functional independence of the derivatives of scalar potentials $\partial_i\Phi_{(r)}$ , the fulfillment of following series of conditions
 is the enough and sufficient condition for the fulfillment of (\ref{nabla_Tik}):
\begin{equation}\label{Eq_S}
\square\Phi_{(r)}+\epsilon^{(r)}_2{m^{(r)}_s}\!\ ^2\Phi_{(r)}=-4\pi\epsilon^{(r)}_1\sigma^{(r)}.
\end{equation}
Thus we obtain the system of equations for the potentials of scalar field of Klein-Gordon kind of equations (accurate within signs) with sources.

\section{Thermodynamic Equilibrium of Plasma in the Gravitational Field}
\label{I.III.1}
\subsection{The Locally Equilibrium Distribution}
At presence of thermodynamic equilibrium it is:
\begin{equation}\label{3.1.1}
\frac{d S}{d\tau} = 0\,.
\end{equation}
Let us first suggest that interactions of all T particles is invariant. Then \Req{3.1.1} can be fulfilled only at fulfillment of the next conditions
 (see \cite{Yubook1}):
\begin{equation}\label{3.1.2}
Z_{fi} - Z_{if} = 0
\end{equation}
in each channel of reactions \Req{GrindEQ__50_}. Equations \Req{3.1.2} are similarities of the functional Boltzmann equations \cite{chern5}. In order to solve them let us make the following
 substitute:
\begin{equation}\label{3.1.3}
F_a = {\rm e}^{- \phi_a}\krugskob{e^{- \phi_a} \mp 1}^{-1} \equiv \krugskob{1 \mp
{\rm e}^{\phi_a}}^{-1}\,,
\end{equation}
as a consequence of which values $Z_{if}$ and
$Z_{fi}$ take form:
\begin{equation}\label{3.1.4}
Z_{if} = \frac{\dsp{\prod_i
e^{-\phi_a}}}{\dsp{\prod_{i,f} (e^{-\phi_a} \mp 1)}}\,;
Z_{fi} = \frac{\dsp{\prod_f e^{-\phi_a}}} {\dsp{\prod_{i,f}
(e^{-\phi_a} \mp 1)}}\,.
\end{equation}
Then after taking the logarithm, equations \Req{3.1.2} take form:
\begin{equation}\label{3.1.5}
\sum\limits_{A=1}^{m} \sum\limits_{\alpha=1}^{\nu_A}
\phi_A(\Pg^{\alpha}_A) = \sum\limits_{B=1}^{m'}
\sum\limits_{\beta=1}^{\nu'_A} \phi'_B(\Pg'^{\beta}_B)\,,
\end{equation}
where these relations must be fulfilled also in each channel of reactions
\Req{GrindEQ__50_}. The unique solution of \Req{3.1.5} at arbitrary particle momentums are the linear functions of the momentums:
\begin{equation}\label{3.1.6}
\phi_A(\Pg^{\alpha}_A) = - \lambda_A(x) + (\xi_A,
\Pg^{\alpha}_A)\,,
\end{equation}
where due to the distribution function invariance $\lambda_A (x)$ are scalars and $\xi^i (x)$ are vectors in the configurational space. Substituting \Req{3.1.6} into equ\-ations \Req{3.1.5} and taking into account the law of generalized momentums' conservation at collisions, we
 obtain as a result of arbitrariness of particle momentums:
\begin{eqnarray}
\label{3.1.7}\xi^i_A (x) = \xi^i (x)\,; \\
\label{3.1.8}
\sum\limits_{A=1}^{N} \nu^k_A \lambda_A = 0\,,
\end{eqnarray}
where $||\nu^k_A||$ is an integer matrix introduced in \cite{Yubook1}.
In consequence of the obvious closure condition of all cycles of the reactions
\begin{equation}\label{2.1.11}
{\rm rank}||\nu^k_A||<N
\end{equation}
equations \Req{3.1.8} always have a nontrivial solution.

{\it Conditions \Req{3.1.7} and \Req{3.1.8} are the conditions of the local thermodynamic equilibrium \\ (LTE); scalars
$\lambda_A(x)$ are called chemical potentials of the statistical system.}

Substituting solutions \Req{3.1.6} into \Req{3.1.3} with an account of \Req{3.1.7}
we find {\it the local equilibrium dis\-tri\-bution functions}:
\begin{equation}\label{3.1.9}
f^0_a (x, \Pg_a) = \figurskob{\exp[ - \lambda_a +
(\xi, \Pg_a)] \mp  1}^{- 1}\,,
\end{equation}
where, as before, the upper sign corresponds to bosons while the lower one - to fermions.

For convergence of the moments of distribution \Req{3.1.9} vector \\ $\xi^i(x)$ should be timelike:
\begin{equation}\label{3.1.10}
\xi^2 \equiv (\xi, \xi) > 0\,.
\end{equation}
Let us with the help of $\xi^i(x)$ introduce a timelike field $v^i(x)$:
\begin{equation}\label{3.1.11}
v^i =
\frac{\xi^i}{\xi}\,; \quad (v,v) = 1\,,
\end{equation}
a local temperature $\theta (x)$ \cite{chern5}:
\begin{equation}\label{3.1.12}
\theta (x) = \xi^{-1}
\end{equation}
and chemical potentials $\mu_a(x)$ in the ordinary nor\-ma\-li\-zation:
\begin{equation}\label{3.1.13}
\mu_a(x) = \theta(x) \lambda_a(x)\,.
\end{equation}
The distribution \Req{3.1.9} can be written in the following form:
\begin{equation}\label{3.1.14}
f^0_a(x,\Pg_a) = \figurskob{\exp
\kvadrskob{\frac{\dsp{- \mu_a + (v,\Pg_a)}}{\theta}} \mp 1}^{ -
1}\,.
\end{equation}

\subsection{The Moments of the Equilibrium Distribution}
Let us calculate the moments of distribution \Req{3.1.9}. Here it is convenient to proceed
to locally Lorentzian reference, time component of which is directed in the line of
 vector $v^i$ and the mass surface equation in which (см. \cite{Ignatev_print})
\begin{equation}\label{P_norm}
(P,P)=m^2_*.
\end{equation}
takes form:
\begin{equation}
\label{3.1.15}
P^2_4 = P^2 + m^2_*\,.
\end{equation}
Then we should proceed to spherical coordinate system in momentum space $P(X)$ and generalize the obtained results covariantly. As a result we obtain the expressions for particle number density vector's components $n^i_a(x)$ and the energy - mo\-mentum tensor of
$a$ component of plasma $\stackunder{a}{T}^{ik}$, \cite{Ignatev3},
\cite{kuza}:
\begin{eqnarray}\label{3.1.17} n^i_a(x) = n_a(x) v^i\,;\\
\label{3.1.18}
\stackunder{a}{T}^{ik}(x) = (\Eps_a + P_a) v^i v^k - P_a g^{ik}.
\end{eqnarray}
Calculating introduced above scalars for each sort of particles, we obtain:
\newcommand{\fo}{\exp\bigl(\frac{- \mu_a + \sqrt{m^2_* + P^2}}{
\theta}\bigr)\mp 1}
\begin{equation}\label{3.1.19}
n_a(x) =
\frac{\rho}{2\pi^2} \dsp{\int\limits_{0}^{\infty}} \frac{ P^2 d P}{\fo}\,;
\end{equation}
\begin{equation}
\label{3.1.20}\Eps_a(x) =
\frac{\rho}{2\pi^2} \dsp{\int\limits_{0}^{\infty}} \frac{\sqrt{m^2_* + P^2} P^2 d P}{\fo}\,; \\
\end{equation}
\begin{equation}
\label{3.1.21} P_a(x) = \frac{\rho}{6\pi^2}
\dsp{\int\limits_{0}^{\infty}}\frac{P^4
d P}{\sqrt{m^2_* + P^2}\fo}\,;
\end{equation}
\begin{eqnarray}
\label{3.1.21a} \sigma^{(r)}_a(x) = \frac{\rho m_*q^{(r)}_a}{2\pi^2}\times\nonumber\\
\dsp{\int\limits_{0}^{\infty}} \frac{P^2
d P}{\sqrt{m^2_* + P^2}\fo}\,.
\end{eqnarray}
In such case:
\begin{equation}\label{sum}
\Eps=\sum\limits_a \Eps_a; \; P=\sum\limits_a P_a; \; \sigma^{(r)}=\sum\limits_a \sigma^{(r)}_a.
\end{equation}

Let us notice that unit vector in the direction of particle number density vector is called a {\it kinematic medium velocity} and unit timelike eigenvector of particle energy - momentum tensor is called a {\it dynamic medium velocity}. Eigenvalue of energy - momentum tensor corresponding to this vector is called a {\it medium energy density} (see e.g. \cite{Sing}).
Thus, at conditions of local thermodynamic equilibrium, particle kinematic velocity coincides with their dy\-namic velocity and is equal to $v^i$.

\subsection{Symmetries of Thermodynamic Equilibrium}

A chemical potential of massless particles having zero fundamental charges at conditions of local thermodynamic equilibrium is equal to zero. This conclusion follows from the fact that numbers $\nu^k_A$ of such particles participating in reactions  \Req{3.1.8} can be absolutely arbitrary. Then from the fact of existence of the reaction of particles
 and particles annihilation it follows the next well-known relation \cite{Landau_Stat}:
\begin{equation}\label{3.1.24}
\anti{\mu}_a = - \mu_a\,.
\end{equation}
Let then exists $N$-plet of scalar fields:
\begin{equation}\label{n-plet}
\mathbf{\Phi}=\{\Phi_1,\Phi_2,\ldots,\Phi_N\}.
\end{equation}
Let us find out how the macroscopic scalar densities (\ref{3.1.19}) -- (\ref{3.1.21a}) are transformed relative to transformation:
\begin{equation}\label{trans_F}
\mathbf{\Phi}:\; \mathbf{\Phi}\rightarrow -\mathbf{\Phi}.
\end{equation}
At transformations $\mathbf{\Phi}$ (\ref{trans_F}) effective particle mass (см. \cite{Ignatev_print})
\begin{equation}\label{m_*}
m^a_*=m^a_0+\sum\limits_r q^{(r)}_a\Phi_r
\end{equation}
is transformed by the law:
\begin{equation}\label{trans_m}
m_*(-\mathbf{\Phi})=m_0-\sum\limits_r  q^{(r)}_a\Phi_{(r)}=2m_0-m_*(\mathbf{\Phi}).
\end{equation}

Therefore at transformations $\mathbf{\Phi}$ (\ref{trans_F}) effective masses of particles and
antiparticles are tied by the relation:
\begin{equation}\label{m_anti}
m_*(-\mathbf{\Phi})=\anti{m}_*(\mathbf{\Phi}).
\end{equation}

Thus, we obtain the transformation laws of scalar densities (\ref{3.1.19}) -- (\ref{3.1.21a}) relative to
transformation $\mathbf{\Phi}$  (\ref{trans_F}):
\begin{eqnarray}
\label{n,e}
n_a(-\mathbf{\Phi})=\anti{n}_a(\mathbf{\Phi}); & \Eps_a(-\mathbf{\Phi})=\anti{\Eps}_a(\mathbf{\Phi});\\
\label{p,s}
P_a(-\mathbf{\Phi})=\anti{P}_a(\mathbf{\Phi}); & \sigma^{(r)}_a(-\mathbf{\Phi})=\anti{\sigma}^{(r)}_a(\mathbf{\Phi}),
\end{eqnarray}
i.e. {\it macroscopic scalar densities are invariant relative to transformation (\ref{trans_F})}.

Let us consider now the transformation of charge conjugation:
\begin{equation}\label{qtrans}
\mathbf{Q}:\; q^{(r)}_a\longleftrightarrow -q^{(r)}_a,\Rightarrow \mathbf{q} \longleftrightarrow -\mathbf{q},
\end{equation}
where $\mathbf{q}$ is a charge matrix of $n\times N$ size:
$$\mathbf{q}=\left(\begin{array}{cccc}
q^{(1)}_1 & q^{(1)}_2 & \cdots & q^{(1)}_n\\
q^{(2)}_1 & q^{(2)}_2 & \cdots & q^{(2)}_n\\
\cdots & \cdots & \cdots & \cdots \\
q^{(N)}_1 & q^{(N)}_2 & \cdots & q^{(N)}_n\\
\end{array}\right).
$$

Let us investigate scalar densities (\ref{3.1.19}) -- (\ref{3.1.21a}) transformation law at charge transformation
 $\mathbf{Q}$ (\ref{qtrans}). We obtain the effective mass transformation law:
\begin{equation}\label{m_q_trans}
m_*(-\mathbf{q})= 2m_0-m_*(\mathbf{q})
\end{equation}
and similarly to (\ref{m_anti}):
\begin{equation}\label{m_anti_q}
m_*(-\mathbf{q})=- \anti{m}_*(\mathbf{q}),
\end{equation}

Let us consider now particles and antiparticles in thermodynamic equilibrium. In this case relation (\ref{3.1.24}) between chemical potentials of particles and antiparticles should be taken into account as well as the fact that
in sums over particle sorts both particles and antiparticles occur in same positions.
Then we can find the summary densities' tran\-s\-for\-mation laws:
\begin{eqnarray}\label{n_anti}
n_a(-\mathbf{q},-\mu)=& \anti{n}_a(\mathbf{q},\mu);\\  \label{P_anti}P_a(-\mathbf{q},-\mu)=& \anti{P}_a(\mathbf{q},\mu)\\
\label{E_anti}
\Eps_a(-\mathbf{q},-\mu)=& \anti{\Eps}_a(\mathbf{q},\mu);\\
\label{s_anti} \sigma^{(r)}_a(-\mathbf{q},-\mu)=& \anti{\sigma}^{(r)}_a(\mathbf{q},\mu),
\end{eqnarray}
i.e. scalar densities are invariant also with respect to charge conjugation $\mathbf{Q}$ subject to change $\mu\to\anti{\mu}=-\mu$.
Thus, at least {\it in conditions of local thermodynamic equilibrium, the kinetic theory of  statistical systems with scalar particle interactions, if expanded to range of negative masses, is invariant with respect to scalar-charge conjugation}.

Let us then notice that the formula for the effective mass is invariant with respect to product of transformations (\ref{trans_F}) и
(\ref{qtrans}) $\mathbf{Q}\times \mathbf{\Phi}$:
\begin{equation}\label{qFtrans}
m_*(-\mathbf{q},-\mathbf{\Phi})=m_*(\mathbf{q},\mathbf{\Phi}).
\end{equation}
Therefore we find the following formulas for scalar densities tran\-s\-for\-mation with respect to composition of
 transformations $\mathbf{Q}\times \mathbf{\Phi}$ :
\begin{eqnarray}\label{n_q_F}
n_a(-\mathbf{q},-\mathbf{\Phi})=& n_a(\mathbf{q},\mathbf{\Phi});\\  \label{P_q_F}P_a(-\mathbf{q},-\mathbf{\Phi})=& P_a(\mathbf{q},\mathbf{\Phi})\\
\label{E_q_F}
\Eps_a(-\mathbf{q},-\mathbf{\Phi})=& \Eps_a(\mathbf{q},\mathbf{\Phi});\\
\label{s_q_F} \sigma^{(r)}_a(-\mathbf{q},-\mathbf{\Phi})=& -\sigma^{(r)}_a(\mathbf{q},\mathbf{\Phi}),
\end{eqnarray}
Thus, scalar charge density change its sign.

This last property ensures internal consistency of the theory. Indeed, exposing scalar field equations (\ref{Eq_S}) to transformation $\mathbf{Q}\times \mathbf{\Phi}$ with an account of (\ref{s_q_F}) and linearity of the left part of equations (\ref{Eq_S}), we find the original equations. Thus, scalar field equations (\ref{Eq_S}) are also invatiant with respect to transformations $\mathbf{Q}\times \mathbf{\Phi}$. The similar case exists also in Maxwell theory as well as in theory of classic gravitational field.

\subsection{A Case of Zero Bare Mass}
As is shown in \cite{Ignatev_print}, when selecting effective mass function in form
\begin{equation}\label{m0=0}
m_0=0\Rightarrow m_*=\sum\limits_r q^{(r)}_a\Phi_{(r)},
\end{equation}
i.e. if bare mass is selected to be equal to zero, the theory becomes symmetric with respect to any of transformations (\ref{trans_F}) and
(\ref{qtrans}) at a level of microscopic dynamics:
\begin{equation}\label{qandF}
m_*(-\mathbf{q})=- m_*(\mathbf{q}),\quad m_*(-\mathbf{\Phi})=- m_*(\mathbf{\Phi}).
\end{equation}
Thus, the following transformation laws take place:
\begin{eqnarray}\label{n_q_m0}
n_a(-\mathbf{q})= n_a(\mathbf{q});&  P_a(-\mathbf{q})= P_a(\mathbf{q})\\
\label{E_q_m0}
\Eps_a(-\mathbf{q})= \Eps_a(\mathbf{q}); & \sigma^{(r)}_a(-\mathbf{q})=\sigma^{(r)}_a(\mathbf{q}),
\end{eqnarray}
as well as:
\begin{eqnarray}\label{n_F_m0}
n_a(-\mathbf{\Phi})= n_a(\mathbf{\Phi});&  P_a(-\mathbf{\Phi})= P_a(\mathbf{\Phi})\\
\label{E_F_m0}
\Eps_a(-\mathbf{\Phi})= \Eps_a(\mathbf{\Phi}); & \sigma^{(r)}_a(-\mathbf{\Phi})=-\sigma^{(r)}_a(\mathbf{\Phi}).
\end{eqnarray}
Thus, scalar charge density changes its sign with respect to
transformation $\mathbf{\Phi}$.

\subsection{T-Noninvariant Interactions}
Let us now consider T-noninvariant interactions assuming however that particle distributions remains equilibrium i.e. \Req{3.1.9}. But then functional relations \Req{3.1.2} are identically fulfilled and then entropy rate of change is equal to zero i.e. we again obtain the LTE condition \Req{3.1.1}. Thus, system entropy is always conserved if particle distribution is locally equilibrium.

Let us now return to transport equations \Req{GrindEQ__54_}. In conditions of LTE due to \Req{3.1.3} these equations take simpler form:
\begin{eqnarray}\label{3.1.26}
\nabla_i \sum\limits_{A}^{} \intx{\Pg} \psi f p^i d\pi -
\sum\limits_{A}^{} \intx{\Pg} f \Puass{\Psi} d \pi =\nonumber\\
 - \sum\limits_{k}^{} \intx{\Pg}\delta^{(4)}(\Pg_F - \Pg_I)\times \nonumber\\  \sum\limits_{A}^{} \nu^k_A \psi_A
Z_{if} (W_{if} - W_{fi}) \prod_{i,f} d \pi\,.
\end{eqnarray}
Putting, in particular, $\psi_A = \delta^a_A$, from \Req{3.1.26} we find:
\begin{eqnarray}\label{3.1.27}
\nabla_i n^i_a = - \sum\limits_{k}^{} \nu^k_A
\intx{\Pg} \delta^{(4)} (\Pg_F - \Pg_I)\times\nonumber\\
Z_{if} (W_{if} - W_{fi})\prod_{i,f} d \pi\,.
\end{eqnarray}
Integrating the right part of \Req{3.1.27} over final states of particles we have:
$$
\sum \nu^k_A \int \prod_f (1 \pm f) (W_{if} - W_{fi}) d \pi_f\,.
$$
Due to unitary property of the $S$-matrix and the optical theorem
(see \cite{Yubook1}) this integral is equal to zero. Therefor in conditions of LTE the conservation law of each particle sort
 conservation takes place:
\begin{equation}\label{3.1.28}
\nabla_i n^i_a = 0 \Leftrightarrow N_a = \Const\,.
\end{equation}
In \eqref{3.1.28} by $n_a$ we should mean particles' and anti\-par\-ticles' density difference
:
\begin{equation}\label{dn_a}
N^+_a-N^-_a=\Const.
\end{equation}

\subsection{A Global Thermodynamic Equilibrium} \label{I.III.3}
The theory of global thermodynamic equilibrium is generalized trivially to the case of possibility of negative effective masses. Therefore we give here just basic relations referring reader to Author's earlier works \cite{Ignatev4,kuza}. In case when distribution functions \Req{3.1.9} (or \Req{3.1.14}) are exact solutions of the kinetic equations the statistical system lies in strict {\it global thermodynamic equilibrium}. In conditions of global thermodynamic equilibrium the laws of each particle sort conservation \Req{3.1.28}a re filfilled and system entropy is strictly constant $S = \Const$. To find the conditions of global thermodynamic equilibrium in case of T-invariant interactions let us substitute solutions \Req{3.1.9} in the kinetic equations. Since the integral of T-invariant interactions turns to zero on locally equilibrium
distributions, let us reduce the kinetic equations to the form:
\begin{equation}\label{3.2.1}
\kvadrskob{H_a, \phi_a} = 0\,,
\end{equation}
where it is required to substitute the expression for $\phi_a$ из (\ref{3.1.6}). With an account of relation (\ref{HPsi}), we obtain:
\begin{equation}\label{Hphi}
[H_a,\phi_a]= \frac{1}{m_*}\left(P^iP^k\xi_{(i,k)}-P^i\lambda_a+\partial_i m_* \xi^i\right)=0.
\end{equation}
Thus, to ensure GTE there should exist a linear integral of motion where $\xi^i$ is a timelike vector. Taking into account the arbitrariness of the mo\-mentum vector, we obtain at
 $m_*\not\equiv 0$ the enough and sufficient conditions of GTE existence:
\begin{eqnarray}
\label{GTE_g}
\Lee{\xi}g_{ik}=0;\\
\label{GTE_m}
\Lee{\xi}m_*=0;\\
\label{GTE_l}
\lambda_a=\Const.
\end{eqnarray}
As a result of effective mass definition (\ref{m0=0}) and functional independence of scalar fields we can find
more strict conditions of GTE from (\ref{GTE_m}):
\begin{equation}\label{LPhi}
\Lee{\xi}\Phi_{(r)}=0,\quad (r=\overline{1,N}).
\end{equation}

Next, since all moments of the equilibrium dis\-tri\-bution function are determined via scalars $\xi^2$,
$\lambda_a$, $\Phi_{(r)}$ and tensors $\xi^i$, $g^{ik}$,
$\xi^i \xi^k$, ..., then dis\-tri\-bution function moments' conservation laws \cite{Ignatev3} are also fulfilled:
\begin{equation}\label{3.2.9}
\Lee{\xi} n^i_a =0\,;
\end{equation}
\begin{equation}\label{3.2.11}
\Lee{\xi} \stackunder{a}{T}^{ik} = 0
\end{equation}
etc.  As a result of these equations in the direction of
$\xi^i$ there are conserved the components of Riemannian tensor, Ricci tensor and Einstein tensor:
\begin{equation}\label{3.2.13}
\Lee{\xi} R_{ijkl} = 0\,; \quad \Lee{\xi}
R_{ij} = 0\,; \quad \Lee{\xi} G_{ij} = 0\,.
\end{equation}
Therefor in consequence of the Einstein equations the following relations should fulfill:
\begin{equation}\label{3.2.14}
\Lee{\xi} T_{ij} = 0\,.
\end{equation}
\section{The Conclusion}
In this article we extended the kinetic theory of interacting particles to the case of arbitrary number of scalar fields, having removed the artificial sug\-ges\-tion about nonnegativity of particle effective mass.
For correct generalization of the kinetic theory to negative effective masses of particles we were required to review the series of key points of this theory. In the end we have obtained a natural generalization of the macroscopic theory of substance interaction with scalar fields. This generalization of the theory is free of any problems in case of negative effective masses of particles. It should be noted that in case of zero bare mass the theory becomes completely invariant with respect to charge con\-ju\-gation. Let us also highlight that he sign of particle effective mass function does not impact on positive definition of particle number density, pressure and energy density but only it has an impact on the sign of scalar charge density. In next article we are going to apply the obtained results to certain cosmological and astrophysical problems.

\end{document}